\documentclass[aps,prl,reprint,amsmath,amssymb,groupedaddress,longbibliography]{revtex4-2}

\usepackage{graphicx}
\usepackage{dcolumn}
\usepackage{bm}
\usepackage[hidelinks]{hyperref}

\begin{document}
\title{Synaptic bundle theory for spike-driven sensor-motor system:
  More than eight independent synaptic bundles collapse reward-STDP learning}

\author{Takeshi Kobayashi, Shogo Yonekura, Yasuo Kuniyoshi}

\affiliation{Laboratory for Intelligent Systems and Informatics, Department of Mechano-Informatics, Graduate School of Information Science and Technology, The University of Tokyo, Bunkyo-ku, Tokyo 113-8656, Japan}

\date{\today}

\begin{abstract}
  Neuronal spikes directly drive muscles and endow animals with agile movements, but applying the
  spike-based control signals to actuators in artificial sensor-motor systems inevitably causes a collapse of learning.
  We developed a system that can vary \emph{the number of independent synaptic bundles} in sensor-to-motor connections.
  This paper demonstrates the following four findings:
  (i) Learning collapses once the number of motor neurons or \emph{the number of independent synaptic bundles} exceeds a critical limit.
  (ii) The probability of learning failure is increased by a smaller number of motor neurons, while (iii) if learning succeeds, a smaller number of motor neurons leads to faster learning.
  (iv) The number of weight updates that move in the opposite direction of the optimal weight can quantitatively explain these results.
  The functions of spikes remain largely unknown. Identifying the parameter range in which learning systems using spikes can be constructed will make it possible to study the functions of spikes that were previously inaccessible due to the difficulty of learning.
\end{abstract}

\maketitle

Animals achieve high energy efficiency and adaptive behavior through information transmission via spikes---brief action potentials in neurons. Spiking neural networks (SNNs) explicitly incorporate these spikes as computational primitives \cite{maass1997networks-026}. SNNs have been used to model peripheral sensory systems \cite{wiesenfeld1995stochastic-50a,collins1995stochastic-bbf}, cortical microcircuits model \cite{maass2002realtime-093}, and neural implementations of Bayesian inference \cite{buesing2011neural-3b3,habenschuss2013stochastic-002,habenschuss2013emergence-596,nessler2013bayesian-c3d}. From an engineering perspective, they offer high energy efficiency \cite{gonzalez2024spinnaker2}, the ability to escape local optima \cite{jonke2016solving-e64} and emergence of instant adaptability \cite{yonekura2020spikeinduced-1f3}.
SNNs have also been applied to sensor-motor learning in robotics \nocite{how_spike_to_continuous} \cite{tieck2019learning-2c5,rueckert2016recurrent-07d,juarezlora2022rstdp-f04,chadderdon2012reinforcement-7b9,zanatta2024exploring-9b3,jiang2024spike-bb1,depasquale2023centrality-c70}. \nocite{spike_based_discontinuous_but_not_stdp}

It has been suggested that sensor-to-motor learning in animals is supported by reward-modulated spike timing-dependent plasticity (R-STDP) \cite{legenstein2010rewardmodulated-a04}. Intuitively, if autonomous robots represent sensory inputs and motor outputs as spikes and learn with R-STDP, they could achieve basic sensor-to-motor tasks like stabilization. However, this letter shows that simply applying R-STDP to sensor-to-motor learning is not sufficient to acquire adequate spiking activity of motor neurons.

\begin{figure}[h]
  \centering
  \includegraphics[width=\columnwidth]{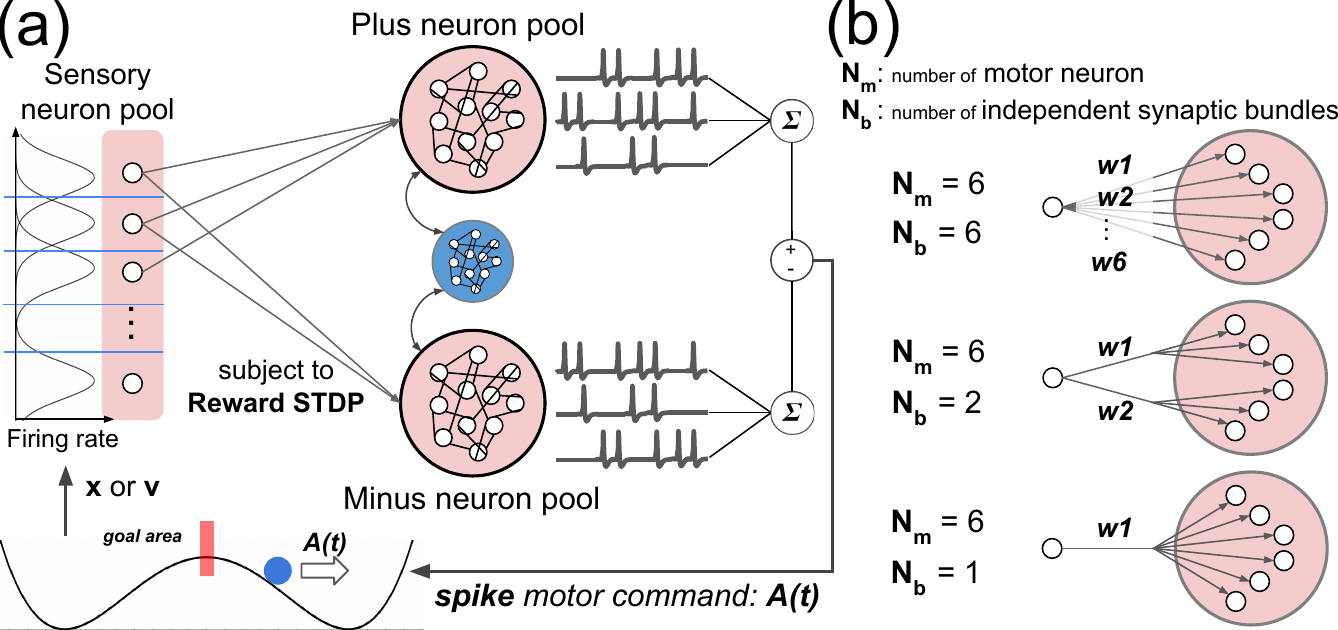}
  \vspace{-20pt}
  \caption{(a) Double-well potential task and network architecture.
    $m=0.3$ and $\gamma=0.5$ is used. The SNN controller consists of separate sensory and motor neuron pools.
    The position $x$ and the velocity $v$ are coded by two distinct sensory pools (only one is shown).
    Each sensory pool contains 30 neurons with tuning curves
    $f_i(s)=c \exp\!\bigl(k(\cos(s-s_i)-1)\bigr)$, where $c=40$, $k=12.5$, and $s_i$ spans $[-1.5,1.5]$ in 30 equal steps. Motor neurons are divided into pools that generate positive and negative thrust. These pools form a winner-take-all circuit via lateral inhibition. The number of neurons in inhibitory population is $2 N_m / 4$. The synaptic weights from positive or negative neuron population to inhibitory population $W_{EI}$ is $75.0 / N_m$, from inhibitory population to positive or negative neuron population $W_{IE}$ is $150.0 / (2 N_m / 4)$. The synaptic weights of recurrent connection within positive or negative neuron population $W_{EE}$ is $8.25 / N_m$. The synaptic weights within winner-take-all circuit is not subjected to Reward STDP learning. The signed sum of the mean PSPs from the positive and negative pools yields the spike motor command $A(t)$ (Eq.~\ref{eq:psp_readout}). Sensor-motor synapses are trained by reward-modulated STDP. (b) Learning performance is systematically examined by varying $N_b$, \emph{the number of independent synaptic bundles}, which is defined as the number of distinct weight values that can be assigned to $N_m$ synapses from a single sensory neuron to a motor pool.
    When $N_m = 6$ and $N_b = 1$, all six synapses share one weight. When $N_b = 2$, the synapses form two groups of three with identical weights. When $N_b = 6$, each synapse has an independent weight.}
  \vspace{-12pt}
  \label{fig:double_well_potential}
\end{figure}

Typically, synaptic weights are updated independently during learning (Fig.~\ref{fig:double_well_potential}(b), top)\cite{tieck2019learning-2c5,rueckert2016recurrent-07d,juarezlora2022rstdp-f04,chadderdon2012reinforcement-7b9}. In this approach, we incorporate a weight-sharing constraint into the learning rule, enabling partial or full coupling of the weights (Fig.~\ref{fig:double_well_potential}(b), middle/bottom). We define the number of synaptic parameters between one sensory neuron and multiple motor neurons as $N_b$, \emph{the number of independent synaptic bundles}. We then systematically examine how the combination of $N_b$ and $N_m$, the number of motor neurons, affects learning performance.
Here, $N_m$ determines the output variance and $N_b$ sets the number of learnable parameters. Experiments revealed a parameter region in which spike-based R-STDP learning is successful: $1 \le N_b \le 8$ and $6 \le N_m \le 20$.
This results show that spike learning actually requires shared synaptic weights (small $N_b$). This finding was revealed only because $N_b$ could be varied in our setup. The results also suggest that the absence of $N_m \to \infty$ and $N_b \to \infty$ organisms in nature may be due not only to energetic limitations, but also to an inherent upper limit on learnability.

The benchmark problem is the stabilization of a point mass in a double-well potential because this seemingly simple system is formally equivalent to classic unstable equilibrium problems, such as balancing an inverted pendulum and controlling human gait\cite{yonekura2020spikeinduced-1f3}.
The dynamics of the mass $x(t)$ is given by $m\ddot{x} = -\gamma \dot{x} - \frac{dU(x)}{dx} + A(t)$ where $m$ denotes the mass, $\gamma$ denotes the friction coefficient, $A(t)$ denotes the control force generated by the SNN controller which is defined Eq.~(\ref{eq:psp_readout}), and $U(x)$ denotes the potential given by $U(x) = \frac{1}{4}x^4 - \frac{1}{2}x^2$.
The objective of the control is to keep the mass within the interval $[-0.1,\,0.1]$.

The SNN controller architecture is shown in Fig.~\ref{fig:double_well_potential}(a).
For a motor neuron $k$, the instantaneous firing probability at time $t$ is given by
\vspace{-4pt}
\begin{align}
  \rho_k(t) = \exp\bigl(u_k(t)\bigr) \Theta(t-\hat t_k-\tau_{\mathrm{ref}})dt
  \label{eq:firing_prob}
\end{align}
where $u_k(t)$ denotes the membrane potential, $\Theta(\cdot)$ is the Heaviside step function, $\hat t_k$ denotes the last spike time, and $\tau_{\mathrm{ref}}$ denotes the refractory period. $u_i(t)$ is given by $u_k(t) = \sum_{i \in \mathrm{pre_k}} I_i(t) + b_k$ where $\mathrm{pre}_k$ denotes the set of presynaptic neurons projecting to neuron $k$, $b_k$ denotes the bias potential. We set $\tau_{\mathrm{ref}}$ to $5$ms and $2$ms and set $b_k$ to $0.0$ and $-1.0$ for excitatory and inhibitory neurons, respectively.
$I_i(t)$ denotes the post synaptic potentials (PSPs) given by 
\begin{align}
  I_i(t) = w_{ik}(t) \bigl(S_i(t) * \epsilon(t)\bigr)
  \label{eq:psp_current}
\end{align}
where $*$ denotes convolution operator, $w_{ik}(t)$ denotes the synaptic weight from neuron $i$ to $k$, $S_i(t)$ is the spike train defined as a sum of Dirac delta functions $S_i(t) = \sum_{l} \delta(t - t_i^{(l)})$ where $t_i^{(l)}$ denotes the $l$-th spike time of neuron $i$. $\epsilon(t)$ is PSP kernel defined as $\epsilon(t) = \Theta(t)\,\bigl(\exp(\frac{-t}{\tau_m}) - \exp(\frac{-t}{\tau_r})\bigr)$ with time constant $\tau_m=20$ms and $\tau_r=2$ms for excitatory neurons and $\tau_m=50$ms and $\tau_r=5$ms for inhibitory neurons.

The synaptic weights $w_i(t)$ between the sensor and motor neurons are updated by an online reinforcement learning rule of the STDP type \cite{nessler2013bayesian-c3d, kappel2018dynamic-1d3}. To represent the presence and magnitude of a connection with a single variable, the synaptic parameter $\theta_i$ is introduced, and the corresponding synaptic weight $w_i$ is defined as follows:

\vspace{-12pt}
\begin{align}
  w_i(t) = \left\{
  \begin{array}{ll}
    \exp(\theta_i(t) - \theta_0) & \text{if } \theta_i(t) > 0 \\
    0                            & \text{otherwise}
  \end{array}
  \right.
\end{align}
where $\theta_0$ denotes offset parameter. We set $\theta_0$ to 3.0.

The update rule for $\theta_i$ is given by
\vspace{-4pt}
\begin{align}
  d\theta_i(t) & = \eta g_i(t) dt + \sqrt{2 \eta T} dW
  \label{eq:theta_update}
\end{align}
where $\eta$ denotes the learning rate, $T$ denotes the temperature parameter, $g_i(t)$ denotes the local estimate of the reward gradient, and $dW$ denotes the Wiener increment. We set $\eta$ to $1.5\times10^{-4}$ and $T$ to $0.1$.
To compute $g_i(t)$, we introduce an eligibility trace $e_i(t)$ that can store the recent history of pre- and postsynaptic spike timing. The method for varying $N_b$ (Fig.~\ref{fig:double_well_potential}(b)) is implemented by replacing the usual one-to-one synaptic connections with virtual one-to-many synapses whose weights are shared. More precisely, each synaptic bundle receives input from a single sensory neuron and projects to $N_{\text{post}}$ motor neurons, where $N_{\text{post}} = N_m / N_b$. Synaptic connections are established from one sensory neuron to $N_m$ motor neurons using $N_b$ synaptic bundles. This $N_b$-bundle connectivity pattern is applied for every sensory neuron to both motor-neuron pools. In this study, we restrict $N_m$ to values divisible by $N_b$. To enable STDP-style learning with such one-to-$N_{\text{post}}$ connections, eligibility trace is approximated by averaging the contributions of postsynaptic neurons. Therefore, the dynamics of $e_i(t)$ and $g_i(t)$ are given by
\vspace{-4pt}
\begin{align}
  \frac{d e_i(t)}{dt} & = - \frac{e_i(t)}{\tau_e} + \frac{I_i(t)}{N_{\text{post}}} \sum_{j=1}^{N_{\text{post}}} (\delta(t - \hat t_j) - \rho_\text{j}(t))
  \label{eq:elgibility_trace}                                                                                                                             \\
  \setlength{\jot}{2pt}
  \frac{d g_i(t)}{dt} & = - \frac{g_i(t)}{\tau_g} + r(t) e_i(t)
  \label{eq:reward_gradient}
\end{align}
where $\tau_e$ and $\tau_g$ denote the time constants for eligibility trace and reward gradient, respectively. We set $\tau_e$ to $1.9\,\mathrm{s}$ and
$\tau_g$ to $50\,\mathrm{s}$. $r(t)$ is the reward signal which becomes larger as the point mass approaches the goal area. $e_i(t)$ well-approximates the spiking-timing synaptic depression-potentiation dynamics \cite{kappel2018dynamic-1d3,fremaux2010functional-d28}.
When presynaptic spikes precede a postsynaptic spike ($t_{\mathrm{post}} - t_{\mathrm{pre}} > 0$), $e_i(t)$ is increased.
Conversely, when presynaptic spikes occur after the postsynaptic spike ($t_{\mathrm{post}} - t_{\mathrm{pre}} < 0$), $e_i(t)$ is decreased. At $t = t_\mathrm{post}$, $I_i(t)$ is typically close to zero. Thus, the positive contribution $\delta(t - \hat t_j)$ does not substantially increase $e_i(t)$. Subsequently, at $t = t_{\mathrm{pre}}$, $I_i(t) > 0$ while $\delta(t - \hat t_j) = 0$, and thus the term $\delta(t - \hat t_j) - \rho_\text{j}(t)$ becomes negative due to $-\rho_\text{j}(t)$, leading to a decrease in $e_i(t)$. Consequently, $e_i(t)$ almost includes the features of STDP \cite{nessler2013bayesian-c3d}.

The spike-based motor command $A(t)$ is obtained from the signed sum of the postsynaptic potentials in the positive and negative motor pools.

\vspace{-8pt}
\begin{align}
  \tau_a \frac{d A(t)}{dt} & = -A(t) \pm \frac{g_a}{N_m} \sum_{k \in \pm\text{motor pool}}^{N_m} \delta(t - \hat t_k) \label{eq:psp_readout}
\end{align}
where $\tau_a$ denotes the time constant of the motor command and $g_a$ denotes the gain parameter. We set $\tau_a$ to $10$ms and $g_a$ to $200$.
In this parameter regime (i.e., $\tau_a = 10\mathrm{ms}$ and $5 \leq N_m \leq 30$), $A(t)$ exhibits spike-based, discontinuous signals\cite{yonekura2020spikeinduced-1f3}.

Learning is carried out online according to Eq.~(\ref{eq:theta_update}).
Five initial positions $x_0$ and five initial velocities $v_0$ are prepared for the candidates, and every combination is trained once per epoch. This results in $25$ episodes per epoch.
The position candidates are $x_0 \in \{-1.0,\,-0.5,\,0.0,\,0.5,\,1.0\}$, and the velocity candidates are $v_0 \in \{-0.35,\,-0.20,\,0.0,\,0.20,\,0.35\}$.
The initial values of $\theta$ are drawn from a uniform distribution of $[3.0,\,5.0]$. Each episode lasts $45.0$ seconds, followed by a $5.0$ second interval to reset the network.
Learning performance is evaluated using a metric referred to as the score. The score is designed such that maintaining $x = 0.0$ and $v = 0.0$ throughout an episode yields the maximum value of $1.0$.

\begin{figure}[h]
  \centering
  \includegraphics[width=0.85\columnwidth]{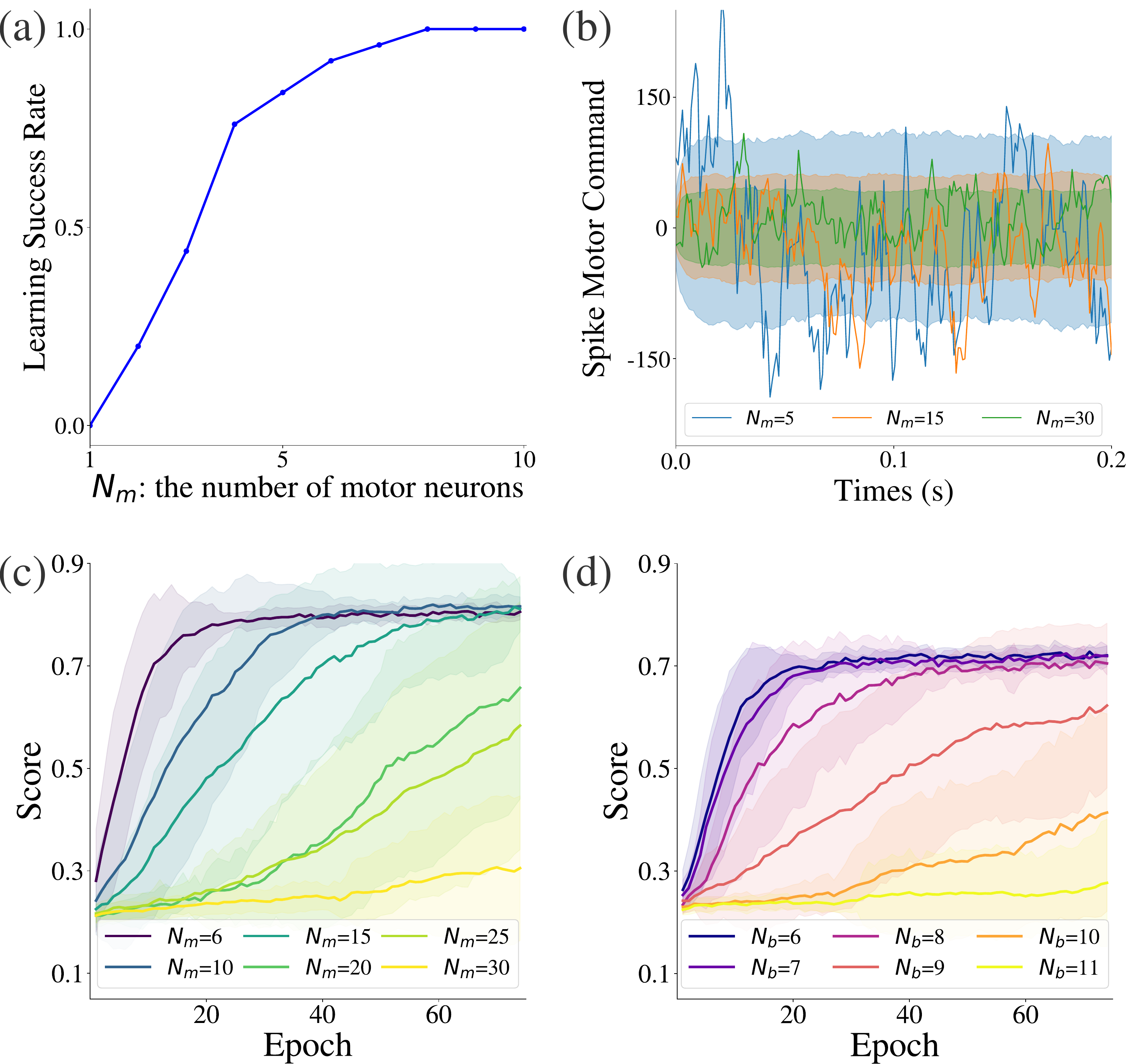}
  \vspace{-8pt}
  \caption{(a) The learning success rate for \(N_m \le 10\). 25 training runs were performed for each value of \(N_m\). A run was considered successful if its final epoch score exceeded 0.65. (b) The effect of \(N_m\) on the amplitude (variance) of the spike motor command.
    For each \(N_m\), 1000 simulations of 0.2 seconds were run, with an average firing probability of 0.15 per neuron. The outputs were generated via Eq.~(\ref{eq:psp_readout}).
    The shaded band indicates the magnitude of the output variance, and the solid line represents a single randomly selected output trace. (c) Learning curves for $N_b = 1$ with $N_m = \{6, 10, 15, 20, 25, 30\}$. 20 training runs were performed for each $N_m$. The solid line shows the mean and the shaded band shows the standard deviation. (d) Learning curves for \(N_b = \{6, 7, 8, 9, 10, 11\}\) with \(N_m = N_b\). 20 training runs were performed for each $N_b$. The solid line shows the mean and the shaded band shows the standard deviation.}
  \label{fig:learning_results}
\end{figure}

To quantify variability across trials, we employ the learning success rate, which is defined as the proportion of trials in which the score exceeds $0.65$. The learning success rate is shown in Fig.~\ref{fig:learning_results}(a) when synaptic weights are shared ($N_b = 1$) and $N_m = \{1, 2, \dots, 9, 10\}$. The success rate is above 90\% for $N_m > 7$. For smaller $N_m$, the success rate declines.
This behavior stems from the dependence of the readout in Eq.~(\ref{eq:psp_readout}) on $N_m$ (Fig.~\ref{fig:learning_results}(b)), indicating that there is an optimal range of $N_m$ for controlling the point mass when combined with the fixed gain $g_a = 200.0$.


\vspace{-12pt}

The learning curves for $N_b = 1$ and $N_m = \{6, 10, 15, 20, 25, 30\}$ are shown in Fig. ~\ref{fig:learning_results}(c). The larger the value of $N_m$, the slower the learning process. For $N_m > 25$, no learning progress is observed because the variance is too small to generate sufficient output to move the point mass (Fig.~\ref{fig:learning_results}(b)).

Finally, Fig.~\ref{fig:learning_results}(d) shows the learning curves for independent synaptic weights ($N_b = N_m$) with $N_b = \{6, 7, \dots, 10, 11\}$. Learning proceeds for $N_b \le 8$, becomes unstable at $N_b = 9$, and fails to progress for $N_b \ge 10$.

To analyze learning performance more precisely, \emph{the number of incorrect transitions} was counted for each synapse at every time step. This is the number of times that the sign of $g_i(t)$ in Eq.~(\ref{eq:reward_gradient}) points in the opposite direction of the correct synaptic weight update.
The target weights were obtained from runs in which learning succeeded with shared weights ($N_b = 1$) and $6 \le N_m \le 20$. Each resulting weight vector was considered the ground truth, and their pairwise cosine similarities were computed. Since every pair scored at least 0.95, $N_b = 1,N_m = 10$ learning weights were selected as the optimal weights.

\begin{figure}[h]
  \centering
  \includegraphics[width=0.90\columnwidth]{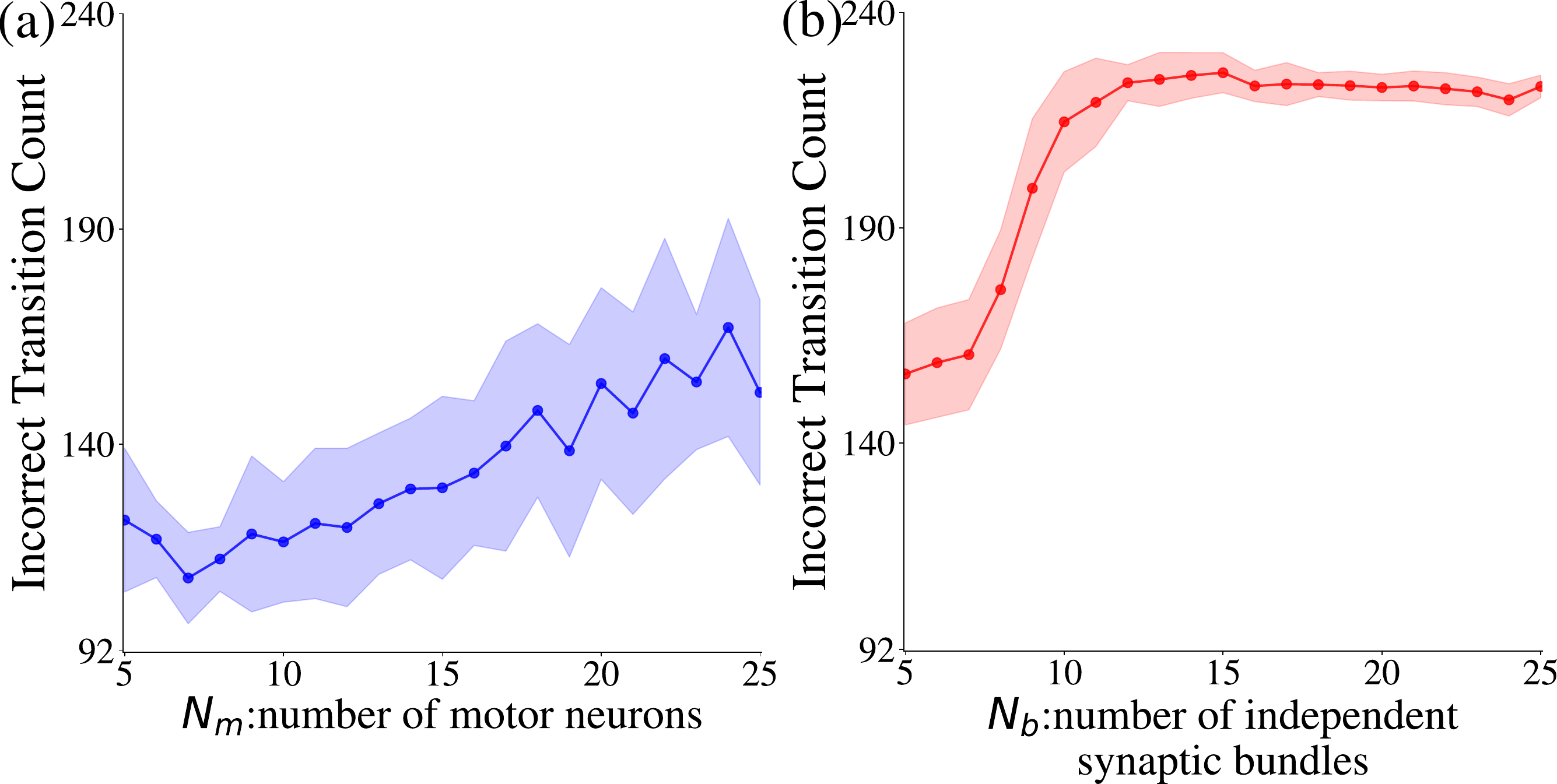}
  \vspace{-12pt}
  \caption{(a) Relationship between the number of incorrect transition counts (vertical axis), which is the number of times a weight update moves in the opposite direction of the correct synaptic weight, and \(N_m\) with \(N_b = 1\). (b) The same metric is plotted against \(N_b\) for networks with \(N_m=N_b\). 15 training runs were performed for each value of \(N_m\) (or \(N_b\)).The solid line shows the mean, and the shaded band shows the standard deviation. The incorrect transition count represents the average number of incorrect transitions per synapse per second.}
  \vspace{-10pt}
  \label{fig:incorrect_count_vs_neurons}
\end{figure}

As shown in Fig.~\ref{fig:incorrect_count_vs_neurons}(a), the number of incorrect transition count increases with $N_m$. Consistent with the earlier observation that learning is faster with smaller $N_m$, runs with fewer motor neurons exhibit fewer incorrect transitions. This confirms that a lower number of incorrect transitions coincides with faster learning.

As shown in Fig.~\ref{fig:incorrect_count_vs_neurons}(b), the number of incorrect transition count increases with $N_b < 10$. The number of incorrect transition counts increases sharply at $N_b$ around $8$. This results accurately shows that smaller $N_b$ are better for learning.

\begin{figure}[h]
  \centering
  \includegraphics[width=0.90\columnwidth]{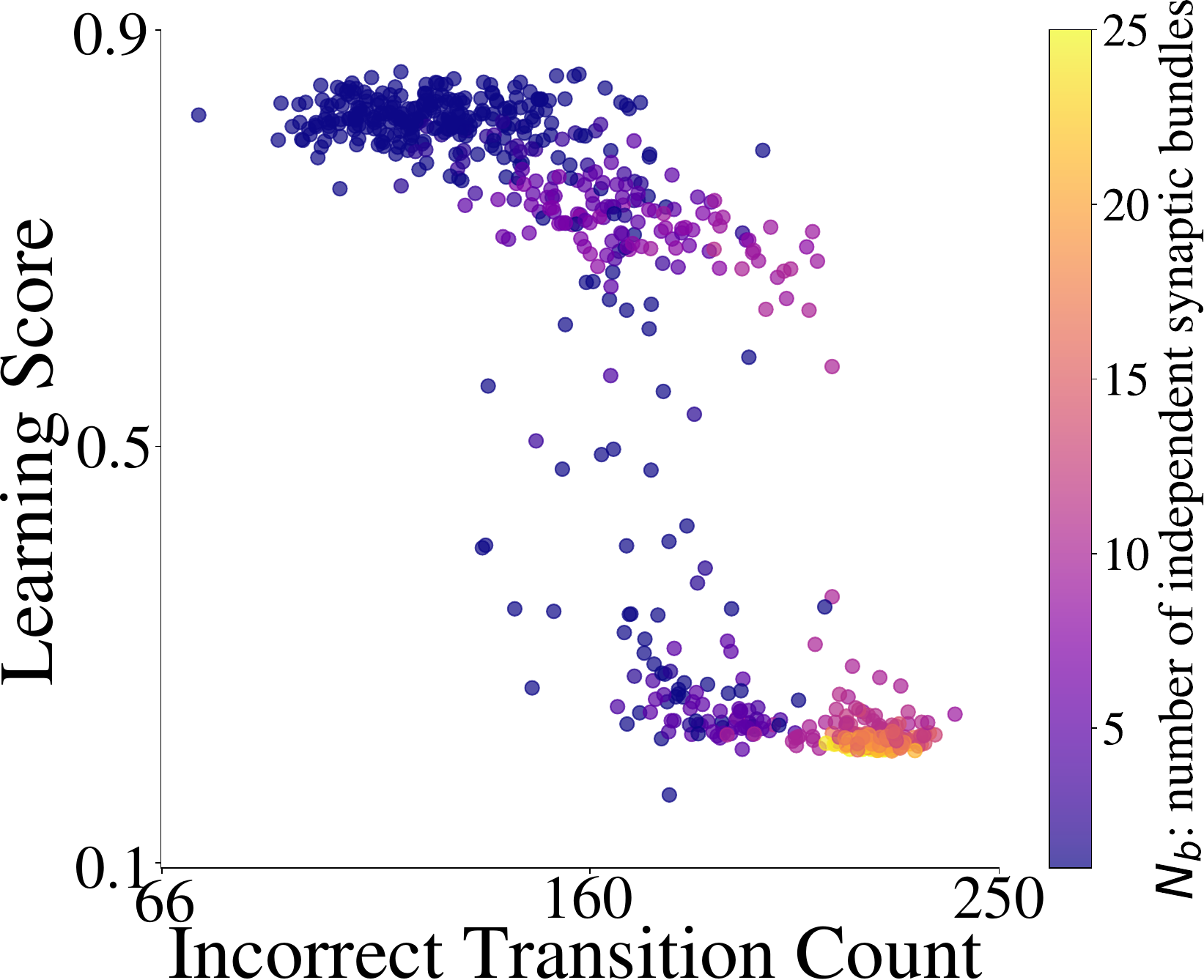}
  \vspace{-8pt}
  \caption{Scatter plot of learning score versus the incorrect transition count which is the number of times a weight update moves in the opposite direction of the optimal synaptic weight. Learning runs were conducted for parameter combinations in which $N_b$ was varied while $5 <= N_m <= 25$.
    Each point represents a single training run. The color of each point corresponds to the value of \(N_b\). For each parameter combination, $15$ training runs were performed. The incorrect transition count represents the average number of incorrect transitions per synapse per second.}
  \label{fig:learning_score_vs_incorrect_count_colored_df}
\end{figure}

Fig.~\ref{fig:learning_score_vs_incorrect_count_colored_df} shows a scatter plot of learning scores versus the incorrect transition counts for $5 \le N_m \le 25$ with varying $N_b$. Learning scores are higher when the incorrect transition count is low. This regime that also corresponds to smaller values of $N_b$.

\vspace{-10pt}

This study represents both sensory inputs and motor outputs entirely with spikes and applies reward modulated STDP reinforcement learning to the simplest sensor-motor task: point mass stabilization in a double well potential. Treating \emph{the number of independent synaptic bundles}, $N_b$, as a controllable parameter allows the number of sensor-to-motor synapses and motor neurons to be varied independently. This enables a comprehensive exploration of the parameter space.

The analysis revealed three key findings:(i) learning remains stable only for neuron counts in the range of $6 \le N_m \le 20$,(ii) smaller $N_m$ leads to faster convergence, and (iii) learning progress is observed only when
$N_b \le 8$. When $N_m<6$, the variance of the postsynaptic potential readout, as given by Eq.~(\ref{eq:psp_readout}), becomes so large that the resulting spike motor command is scaled inappropriately for controlling the point mass (Fig.~\ref{fig:learning_results}(b)).
Conversely, for $N_m>20$ the variance becomes too small to generate control inputs of sufficient magnitude.
Finding (ii) follows because a larger variance at a small $N_m$ enhances the exploratory drive and accelerates learning.
Finding (iii) is counterintuitive. Although more synapses should, in principle, increase representational power, learning actually proceeds better with fewer distinct weights. This effect is attributed to the final motor command being the sum of the positive and negative motor pool activities. Too many independent synaptic bundles cause frequent weight updates in opposing directions, which interferes with convergence. For example, when 10 positive motor neurons and 5 negative motor neurons fire simultaneously, the net force accelerates the point mass in the positive direction. When this movement is rewarded, the synapses that project to the positive pool are reinforced correctly. However, the synapses that project to the negative pool are reinforced incorrectly. Therefore, assigning more than eight distinct synaptic weights ($N_b > 8$) within a single sign motor pool reveals a spatial credit assignment problem.
In fact, as shown in Fig.~\ref{fig:incorrect_count_vs_neurons} and Fig.~\ref{fig:learning_score_vs_incorrect_count_colored_df}, observations (i-iii) are all consistently accounted for by this single measure.

Previous studies have converted the output of motor neurons into a continuous signal \cite{rueckert2016recurrent-07d, zanatta2024exploring-9b3, tieck2019learning-2c5,juarezlora2022rstdp-f04,chadderdon2012reinforcement-7b9,how_spike_to_continuous}.
These techniques seem to have helped avoid the spatial credit assignment problem. In fact, they can reduce reward sensitivity to fluctuations in spike timing. Furthermore, even in studies where control outputs are treated as spike-based discontinuous signals, learning is not based on STDP \cite{yonekura2020spikeinduced-1f3, jiang2024spike-bb1, depasquale2023centrality-c70, spike_based_discontinuous_but_not_stdp}.
However, it has been suggested that the optimal trajectory for rate-based vs. stochastically-spiking control signal is very different \cite{yonekura2020spikeinduced-1f3}. In factor-based approaches \cite{depasquale2023centrality-c70}, SNNs implement trajectories determined at the rate level, where stochasticity and discontinuity are largely suppressed. In contrast, this letter enables learning while preserving stochasticity and discontinuity by introducing \emph{synaptic bundles} and clarifies the conditions under which R-STDP can learn adequate spiking activity.


In biological systems, motor neuron pools are often driven synchronously by a common input—a common drive strategy that has been documented neurophysiologically (e.g.,\cite{luca1994common-995}). In our model, treating $N_b$ as a tunable parameter creates a computational representation of this strategy. The finding that shared input alleviates the spatial credit assignment problem and accelerates learning provides new computational evidence for the functional benefit of common drive—an effect that is evident only when the entire sensor-motor loop is represented by spikes. The collapse of learning when $N_m$ is too small is consistent with the observation of prior research \cite{small_motor_neuron_cannot_convey_common_input, farina2014effective-047}.
Finally, the results of the simulations indicate that learning is most effective when there are only a few neurons and synapses. This finding supports the characteristic resource efficiency of biological systems, in which substantial learning capacity emerges from minimal neural resources.

From the perspective of engineering, demonstrating sensor-to-motor learning while maintaining a spiking representation of motor output allows for the integration of \emph{spike-induced order(SIO)} \cite{yonekura2020spikeinduced-1f3} with reinforcement learning.
\emph{SIO} is an adaptability effect that is expected to emerge in systems with a high degree of freedom.
Although this study was limited to a single degree of freedom, future studies could include scaling to systems with a higher degree of freedom, exploring links to muscle synergies\cite{hug2023common-ec6}, and applying \emph{SIO} more broadly.

This work was supported by JSPS KAKENHI Grant Number JP24KJ0674.

\def\bibsection{}
\bibliography{liter}

\end{document}